\documentclass[conference]{IEEEtran}
\IEEEoverridecommandlockouts

\usepackage{cite}
\usepackage{amsmath,amssymb,amsfonts}
\usepackage{graphicx}
\usepackage{textcomp}
\usepackage{xcolor}
\def\BibTeX{{\rm B\kern-.05em{\sc i\kern-.025em b}\kern-.08em
    T\kern-.1667em\lower.7ex\hbox{E}\kern-.125emX}}

\usepackage{tikz}
\usepackage{filecontents}
\usepackage{float}
\usepackage[english]{babel}
\usepackage{blindtext}
\usepackage{url} 
\usepackage{multicol}
\usepackage{multirow}
\usepackage{comment}
\usepackage{subcaption}
\usepackage{algpseudocode}

\usepackage{algorithm}
\usepackage{bm} 
\algdef{SE}[DOWHILE]{Do}{doWhile}{\algorithmicdo}[1]{\algorithmicwhile\ #1}%
\newcommand{\pvs}{\vspace{-10pt}}
\algtext*{EndIf}
\algtext*{EndFor}
\algtext*{EndWhile}
\algtext*{EndFunction}

\newcommand{\mynewpar}[1]{\pvs~\\\noindent{\bf #1}}

\newcommand{\lam}{\gets}

\renewcommand{\Comment}[1] {\hfill\textit{\textcolor{cerulean}{$\triangleright$~#1}}} 
 
\definecolor{cerulean}{rgb}{0.0, 0.48, 0.65}

\makeatletter

\algnewcommand{\OnlyComment}[1] {\hskip\ALG@thistlm\textit{\textcolor{cerulean}{$\triangleright$~#1}}}
\algnewcommand{\LineIndentComment}[1] {\Statex \hskip\ALG@thistlm\hskip\algorithmicindent\textit{\textcolor{cerulean}{$\triangleright$~#1}}} 
\algnewcommand{\LineComment}[1] {\Statex \hskip\ALG@thistlm\textit{\textcolor{cerulean}{$\triangleright$~#1}}}
\newcommand{\thesys}[1]{Lynceus}
\newcommand{\ts}[1]{\thesys{}}

\begin{document}

\title{Lynceus: Cost-efficient Tuning and Provisioning of Data Analytic Jobs\\
\thanks{
}}

\author{\IEEEauthorblockN{Maria Casimiro$^{1,4}$, Diego Didona$^{2}$, Paolo Romano$^{1}$, Luís Rodrigues$^{1}$, Willy Zwaenepoel$^{3}$, David Garlan$^{4}$}
\IEEEauthorblockA{
\textit{$^{1}$INESC-ID and Instituto Superior Técnico, Universidade de Lisboa}\\
\textit{$^{2}$IBM Research Zurich, }
\textit{$^{3}$University of Sydney, }
\textit{$^{4}$Carnegie Mellon University}}
}

\maketitle
\pagestyle{plain} 
\begin{abstract}

Modern data analytic and machine learning jobs find in the cloud a natural deployment platform to satisfy their notoriously large resource requirements. Yet, to achieve cost efficiency, it is crucial  to identify a deployment configuration that satisfies user-defined QoS constraints (e.g., on execution time), while avoiding unnecessary over-provisioning.

This paper introduces \thesys{}, a new approach for the optimization of cloud-based data analytic jobs that improves over state-of-the-art approaches by enabling significant cost savings both  in terms of the final recommended configuration and of the optimization process used to recommend  configurations. 

Unlike existing solutions, \ts{} optimizes in a joint fashion both the cloud-related 
and the application-level 
parameters. This allows for a reduction of the cost of recommended configurations by up to $3.7\times$ at the 90-th percentile with respect to existing approaches, which treat the optimization of cloud-related and application-level parameters as two independent problems. 

Further, \ts{} reduces the cost of the optimization process (i.e., the cloud cost incurred for testing configurations) by up to $11\times$. Such an improvement is achieved thanks to two mechanisms: 
i) a timeout approach which allows to abort the exploration of configurations that are deemed suboptimal, while still extracting useful information to guide future explorations and to improve its predictive model --- differently from recent works, which either incur the full cost for testing suboptimal configurations or are unable to extract any knowledge from aborted runs;
ii) a \textit{long-sighted and budget-aware} technique that determines which configurations to test by predicting the long-term impact of each exploration --- unlike state-of-the-art approaches for the optimization of cloud jobs, which adopt greedy optimization methods.

\end{abstract}


\section{Introduction}

Many enterprises run data analytic jobs in the cloud, such as training deep neural networks or building recommender systems. This sort of jobs is known to require a very large amount of computational resources and recent studies, e.g.,~\cite{aml19}, have shown that training large AI models can produce five times the lifetime emissions of the average American car (including the manufacturing of the car) and incur cloud costs of up to 3 million USDs.
Further, data analytic jobs are often recurrent, i.e., they execute multiple times on similar datasets, with similar performances~\cite{Alipourfard:2017,Zhang:2016}. 

As such, to reduce operational costs, it is crucial to ensure that jobs are deployed over the cheapest set of cloud resources that complies with user specific constraints, e.g., on job execution time~\cite{Curino:2014,Tumanov:2016}, i.e., it is crucial to optimize the cloud provisioning process so as to avoid over-provisioning. 
Additionally, the efficiency of data analytic jobs can be substantially affected by the correct tuning of a multitude of application-level parameters --- e.g., the hyper-parameters of a machine learning (ML) model can influence its training time~\cite{hypPar}. With hundreds or thousands of possible combinations of cloud platform and job parameters, it is extremely challenging to identify the configuration that minimizes the provisioning cost and meets the target performance constraints.

\mynewpar{Existing solutions and their limitations.} 
State-of-the-art approaches to optimize the deployment of cloud jobs rely on either online or offline learning approaches to find (near) optimal cloud configurations. 

Offline learning techniques require the availability of large training sets, collected by profiling different applications~\cite{Delimitrou:2014,Klimovic:2018,Yadwadkar:2017}, or rely on {\em a priori} knowledge about the internal structure of the job~\cite{Herodotou:2011,Shi:2014,Venkataraman:2016}. These approaches either impose an expensive and time-consuming offline training phase, or require expert domain knowledge to model the performance of a job. We are instead interested in approaches that require no prior knowledge on the target job or other jobs. Therefore in this work we focus on online approaches.

Bayesian Optimization (BO) is a well-established online approach to tackle complex optimization problems~\cite{Brochu:2010,Jones:1998} and has recently emerged as a prominent solution to optimize the execution of data analytic jobs~\cite{Alipourfard:2017,Hsu:2018b,Hsu:2018}. BO approaches profile the job on different configurations iteratively, building at each step a statistical performance model of the job. 
This model is then used to decide the next configuration to try, and ultimately to identify the best configuration for the job.  Unfortunately, BO-based approaches targeting the optimization of cloud jobs suffer from several critical limitations.

\vspace{-10pt}~\\
\noindent\textit{1) Disjoint optimization of cloud and application parameters.} 
Existing BO-based approaches~\cite{Alipourfard:2017,Hsu:2018b,Hsu:2018}
treat the optimization of cloud-related and application-level parameters as independent problems, thus neglecting the existence of important inter-dependencies between cloud and application configurations. 
Note that this limitation also affects existing offline approaches~\cite{Delimitrou:2014,Klimovic:2018,Yadwadkar:2017}.
In Section~\ref{sec:challenges}, we quantify the relevance of adopting a joint, cross-layer optimization approach by means of an experimental study based on three ML jobs, each deployed over 384 (cloud and application level) configurations. Our study shows that  approaches that optimize application and cloud parameters in a disjoint fashion are largely sub-optimal: they find the globally optimal configuration less than 50\% of the times; further, the 90-th percentile of the cost of the recommend solutions is from 1.2$\times$ to 3.7$\times$ larger than the global optimum.

\vspace{-10pt}~\\
\noindent\textit{2) Myopic optimization policy.} 
BO techniques employed by the state-of-the-art solutions~\cite{Alipourfard:2017} demonstrate significant limitations due to their \textit{short-sighted} nature. In fact, at each step of the optimization process, existing solutions profile the configuration that is expected to maximize an immediate reward, such as the Expected Improvement (EI) or the model's accuracy~\cite{Lam:2016,Lam:2017,Gonzalez:2016}. Such greedy approaches are likely to lead to a sub-optimal exploration of the configuration space and require testing a large number of configurations.

\vspace{-10pt}~\\
\noindent\textit{3) Sampling sub-optimal configurations.} 
BO techniques can lead to exploring sub-optimal configurations, especially in the early stages of the optimization process when the model still has very limited information on the job.  Existing BO-based approaches~\cite{Duan:2009,Golovin:2017} address this problem by cancelling the exploration of configurations that are detected to be of lower quality w.r.t. the best configuration identified so far and disregarding any performance metrics obtained during that exploration. However, such an approach suffers from a major drawback: simply cancelling the exploration translates into redundant computational time and money wasted, since the model will not learn from these explorations. This can lead to impoverishing the knowledge of the model on large portions of the configuration space, which has detrimental effects on its accuracy and, as such, on its ability to recommend high quality configurations.

\mynewpar{\ts{}.} This paper presents \ts{} (``Lynx-eyed''), an innovative tool to provision and tune data analytic jobs on the cloud. \ts{} addresses the challenges identified above by combining the following novel features. 

First, \ts{} adopts a cross-layer, holistic approach that optimizes the parameters controlling the cloud deployment as well as the ones defining the application-level configuration \textit{in a joint fashion} and eschews the need for any {\em a priori} knowledge about  the target job. 

Second, \ts{} introduces a novel {\em long-sighted} and {\em budget-aware} optimization method. Differently from existing \textit{greedy} BO-based techniques that maximize a one-step reward, \ts{} plans which configurations to explore by simulating several {\em exploration paths}, i.e., sequences of configurations to sample sequentially. For each path, \ts{} estimates the path's expected exploration cost and the advantages stemming from its exploration (i.e., improvement of model's accuracy and/or of the currently known optimum). While simulating, \ts{} keeps into account predefined constraints both on the configuration's performance and on the cumulative cost of the exploration. By simulating the outcomes of sampling a sequence of configurations, \ts{} determines more cost-effective ways to explore the configuration space. As we will show in Section~\ref{sec:eval:LA}, this technique allows \ts{} to the reduce the cost incurred to optimize the configuration of ML jobs by up to $6.3\times$.

Third, \ts{} uses a new method to cope with the exploration of configurations that turn out to be sub-optimal. Unlike previous approaches~\cite{Duan:2009,Golovin:2017}, \ts{} not only cancels the sampling of sub-optimal configurations to save money and time, but it also exploits the information on \textit{when} such configurations exceeded the cost of the current optimum. \ts{} derives an updated prediction of the actual cost of the job with the sub-optimal configuration. This new prediction leverages the original model's prediction of the execution cost for that configuration, and the time at which sampling was interrupted. This updated prediction is then fed back to the model, which allows for effectively enhancing its knowledge on the regions close to sub-optimal configurations. The use of this technique allows \ts{} to further enhance the cost effectiveness of its own optimization process by up to $1.8\times$ w.r.t. existing approaches that discard information on sub-optimal configurations.

Overall, when combining its innovative features, \ts{} can reduce the 90-th percentile of the cost incurred to find a solution within 10\% from the optimum by up to $11\times$ when compared to state-of-the-art approaches. The experimental results reported in this work were obtained using both existing datasets~\cite{Hsu:2018,Alipourfard:2017}, as well as new datasets obtained by exhaustively deploying three TensorFlow jobs (distributed training of neural networks) over a large 5-dimensional configuration space encompassing 384 configurations.

\mynewpar{Contributions:} We make five main contributions:

\mynewpar{I)} We propose \ts{}, a novel {\em long-sighted and budget-aware} approach to the tuning and provisioning of data analytic jobs, which we will make available as open source;

\mynewpar{II)} We demonstrate the advantages of optimizing both cloud and application parameters jointly;

\mynewpar{III)} We develop a new method for extracting useful information from the partial exploration of sub-optimal configurations;

\mynewpar{IV)} We quantify the gains achievable by  \ts{}  via an extensive experimental study based on 26 diverse jobs; 

\mynewpar{V)} We make available to the systems' community a dataset encompassing three Tensorflow jobs deployed on EC2, each including 384 configurations defined over 5 dimensions~\cite{loura:tf}.

\section{Related Work}
\label{sec:rw}
This section discusses four main kinds of related work: $i)$ systems to tune and provision data analytic jobs; $ii)$ systems to optimize cloud applications; $iii)$ BO approaches to tune generic applications; and $iv)$ variants of the BO approach.

\mynewpar{Optimization of data analytic jobs.}  Elastizer~\cite{Herodotou:2011}, ARIA~\cite{Verma:2011} and MRTuner~\cite{Shi:2014} model the internals of map-reduce jobs and obtain performance models to tune and provision them. Cumulon~\cite{Huang:2013} targets matrix-based big data analysis jobs. Ernest~\cite{Venkataraman:2016} optimizes diverse job types but requires knowledge about the structure of the internal workflow of jobs, e.g., the communication pattern. 
Scout~\cite{Hsu:2018} exploits the availability of historical information on previous cloud jobs to enable transfer learning and navigate through the search space more effectively.
Unlike these approaches, \ts{} needs no {\em a priori} information about the target job or other jobs.
CherryPick~\cite{Alipourfard:2017} and Arrow~\cite{Hsu:2018b}  rely on a greedy BO approach to select the best cloud infrastructure for a job. We discuss the limitations of such an approach in more detail in Section~\ref{sec:challenges} and quantify them in Section~\ref{sec:eval}. In contrast, \ts{} implements a novel \emph{long-sighted} and \emph{budget-aware} BO approach to achieve higher accuracy and better cost-efficiency. In addition, \ts{} tackles jointly the problems of selecting the best cloud infrastructure and optimizing the job's tuning parameters.

\mynewpar{Optimization of cloud applications.} Paragon~\cite{Delimitrou:2013}, Quasar~\cite{Delimitrou:2014}, 
Selecta~\cite{Klimovic:2018} and Paris~\cite{Yadwadkar:2017} optimize the choice of the  infrastructure for cloud applications. These systems employ black-box approaches to performance prediction that rely on the availability of abundant training data on different applications. \ts{} targets scenarios in which such data is not available, and requires running only the target job to infer its performance-cost function.

\mynewpar{BO approaches to tuning systems.}  iTuned~\cite{Duan:2009}, Ottertune~\cite{VanAken:2017},  ProteusTM~\cite{Didona:2016} and Metis~\cite{Lucis:2018} use BO approaches to optimize the tuning parameters of data platforms.
These systems use the traditional BO approach, whose limitations we discuss in detail in Section~\ref{sec:challenges}. Adding to these limitations, ProteusTM requires the availability of previous performance traces of other applications. 
Differently from the previous approaches, BOAT~\cite{Dalibard:2017} extends BO to allow system experts to provide a probabilistic performance model of the target application, so as to speed up the optimization phase. This approach requires expert domain knowledge on the target application. 
Instead, \ts{} embraces a full black-box approach based on a novel long-sighted and budget-aware BO approach and does not require previous performance traces.

\mynewpar{BO with look-ahead.} The ML community  has recently proposed non-greedy BO variants that rely on a \textit{look-ahead} scheme that takes into account future steps in the exploration of the configuration space~\cite{Lam:2016,Lam:2017,Gonzalez:2016}
. These approaches target the optimization of hyper-parameters of ML models, where testing any configuration has a unitary cost, and there is a fixed budget for exploring which is expressed in terms of the number of configurations that can be tested. \ts{} draws from these approaches, but augments them to capture specific idiosyncrasies of cloud environments, and hence to make them suitable in the context of job optimization in the cloud.

In particular, in cloud environments, testing different configurations results in different costs. Note that the cost of exploring a configuration depends on the duration of the job running in that configuration and is not known \textit{a priori}. This is particularly relevant when there is a fixed budget for the optimization process, since it is not known how different explorations will affect the budget available for future explorations. \ts{} copes with this challenge by employing a black-box predictive model to estimate these costs. Further, \ts{} avoids wasting budget in testing suboptimal configurations. 
It achieves this goal via a novel technique to early stop the execution of a job on suboptimal configurations, while still being able to leverage the knowledge attained from the partially completed job to increase the quality of the model.

\begin{figure*}[th!]
  \begin{subfigure}[h]{0.49\textwidth}
       \centering
       \includegraphics[scale = 0.55]{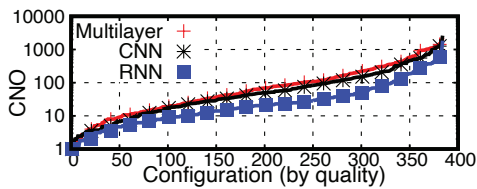}
        \caption{Normalized cost of training 3 Neural Network models with Tensorflow, depending on the configuration (y axis in log scale).}
        \label{fig:challenges:complexity}
    \end{subfigure}
    \hfill
    \hfill
    \begin{subfigure}[h]{0.49\textwidth}
    \centering
       \includegraphics[scale=0.55]{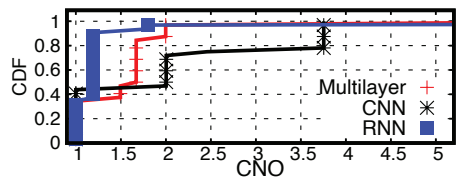}
        \caption{CDF of the normalized cost (w.r.t. the optimum) achievable by optimizing hyper-parameters and cloud configuration separately.}
        \label{fig:challenges:joint}
    \end{subfigure}
\caption{Challenges of job optimization. Costs are normalized w.r.t. the cost of the optimal configuration (CNO). a) Few configurations are close-to-optimal and many are highly sub-optimal. b) Disjoint optimization may lead to identifying sub-optimal configurations. 
}\label{fig:challenges}
\end{figure*}

\section{Problem Formalization and Challenges}
\label{sec:challenges}

\ts{} seeks to find the optimal configuration to run a job, while meeting a target performance constraint. A configuration ${x}$ is a tuple $\langle N, H,P\rangle$, where $N$ is the number of VMs rented from the cloud provider, $H$ encodes the hardware characteristics of the VMs (e.g., VCPUs and RAM), and $P$ represents the settings of job-specific tuning parameters (e.g., hyper-parameters of a ML training process). We define the optimal configuration $x^*$ as the one that minimizes the (monetary) cost of executing the job, and that is able to finish it in at most $T_{max}$ time. The cost  of executing the job with configuration $x$, noted $C(x)$, is given by the product of the time taken to run the job with $x$, noted $T(x)$, and the price per unit of time of renting the cloud configuration $x$, noted $U(x)$. We assume a pay-by-the-minute/second pricing scheme, which is typical nowadays in major PaaS infrastructures~\cite{aws:minute,azure:minute,google:minute}.

The optimization process relies on profiling the target job on a subset of configurations. We note such sub-set $S$, and we denote by $C_S$ the cumulative cost of running the job on the configurations in $S$.  Furthermore, we consider an additional constraint on the maximum cost of the profiling phase, $C_S$, which must not exceed a budget $B$. The problem can then be formalized as follows:

\vspace*{-4pt}
$$
\begin{cases} 
min\ C(x)\\ 
s.t.\ T(x) \leq T_{max} \\
s.t.\ C_S \leq B
\end{cases}
$$

Let us now discuss the key challenges to deriving an efficient and practical solution to this   optimization problem.

\vspace{4pt}\mynewpar{I) Lack of {\em a priori} information.}
Given the heterogeneity and complexity of modern data analytic jobs, building  white-box models capable of accurately predicting their performances, independently of their nature, is not a plausible solution in practice. Moreover, gathering data concerning previous optimizations of similar jobs can be too costly or impractical. In order to circumvent these issues, we advocate optimization methods that ensure two key properties.

\mynewpar{\em Black box approach.}
\emph{The optimization process should assume no knowledge about the target job, nor about the cloud infrastructure}. In fact, jobs can have very different structures (e.g., map-reduce vs parameter-server)~\cite{Herodotou:2011,Venkataraman:2016}, and modeling the performance of cloud infrastructures is notoriously a complex task~\cite{Khazaei:2013}. A black-box approach to the job optimization process reduces the modeling effort and is more flexible.

\mynewpar{\em No available data.}
\emph{The optimization process should not rely on a priori performance information about other jobs.} As noted before, some existing techniques to optimize application performance rely on the availability of huge training data~\cite{Delimitrou:2014,Klimovic:2018}.
This information helps in bootstrapping and improving the optimization process. Unfortunately, obtaining large amounts of training data is very costly and time-consuming. Hence, such approach is fit for large service providers  (such as Amazon, Google and Azure), but constrains and is impractical for most cloud users (e.g., small and medium enterprises).

\vspace{4pt}\mynewpar{II) Complexity of the optimization process.}
The plethora of VMs offered by cloud providers, along with the multitude of tunable application-level parameters, generate a search space with hundreds of configurations, with largely different performances. Next, we present empirical data that demonstrates: i) the complexity of the problem at hand, and ii)  the necessity for tuning application and cloud parameters in a joint fashion. 

\mynewpar{\em Very few close-to-optimal configurations.}
\emph{The configuration space includes few close-to-optimal configurations and many highly sub-optimal ones.} To quantify the  complexity of finding the optimal configuration of modern cloud jobs, we measured the performance of training three ML models  (Multilayer, CNN and RNN) with Tensorflow   on AWS, while varying the cloud infrastructure and job hyper-parameters. In total, we considered 384 configurations. More details about these experiments are provided in Section~\ref{sec:eval}.
Figure~\ref{fig:challenges:complexity} shows the cost of running a job in each configuration, normalized w.r.t. the cost of the optimal configuration. Note that the cost of a ``bad'' configuration can be 3 orders of magnitude worse than the cost of the optimal one. In addition, depending on the job, only 5 to 20 configurations have a cost within a factor of two w.r.t. the optimal one. These configurations correspond to 1.5\% and 5\% of the size of the configuration space, respectively.

\mynewpar{\em The need for joint optimization.}
\emph{The cloud infrastructure and the hyper-parameters of the job must be optimized simultaneously.} An approach to simplify the optimization process could be optimizing these two aspects separately, as done by recent systems~\cite{Delimitrou:2013,Delimitrou:2014}.  
This approach, that we call disjoint optimization, first finds the optimal hyper-parameters by profiling the job on a reference cloud infrastructure $c^{\dagger}$, and then finds the optimal cloud settings for the job running with these parameters. 
Disjoint optimization, however, implicitly assumes that the optimal hyper-parameters for $c^{\dagger}$ are also optimal for other cloud settings. In reality, this is usually not the case. Therefore, disjoint optimization is prone to missing the best combination of hyper-parameters and cloud settings. 

To illustrate this fact we apply disjoint optimization to our jobs using all possible configurations as $c^{\dagger}$, and we measure the cost of the configuration that is identified as optimal. We note that in this experiment we assume that both (i) the hyper-parameter optimization on $c^{\dagger}$ and (ii) the subsequent optimization of the cloud configuration are always able to identify the best solution. Hence, these results are an upper bound on the effectiveness of any practical solution using disjoint optimization. Figure~\ref{fig:challenges:joint} reports the CDF of the cost of the configuration identified via this ideal disjoint optimization (using different choices for the initial reference configuration $c^{\dagger}$), normalized w.r.t. the cost of the actual optimal one. For all jobs, disjoint optimization finds the overall optimal configuration less than 50\% of the times. The 50-th percentile of the normalized cost obtained ranges from 1.2 to 2, and the 90-th percentile from 1.2 to 3.7, depending on the job. 

\section{Background on Bayesian Optimization}
\label{sec:BO}

Bayesian Optimization (BO) is a sequential strategy to find the optimum of a function $f$ with an unknown closed form and whose evaluation is expensive~\cite{Brochu:2010,Jones:1998}. 

\mynewpar{BO operations.}
BO builds a statistical model of $f$ iteratively as follows: $(i)$ evaluate $f$ on a set of  initial 
points $x_1 \ldots x_n$ and create a training set $S$ with the pairs $\langle x_i, f(x_i)\rangle$; $(ii)$ build a  model $M$ over $S$ with a regression algorithm; 
$(iii)$ use an {\em acquisition function} to determine 
the next point $x_m$ to evaluate; $(iv)$ evaluate $f(x_m)$, and 
update $S$ and $M$; $(v)$ repeat steps $(ii)$ to $(iv)$ until a stopping criterion is satisfied.
In \ts{}, a point is a configuration,  and the target function to minimize is the cost of running a job.

\begin{figure*}[th!]
  \begin{subfigure}[h]{0.34\textwidth}
       \centering
       \includegraphics[scale = 0.3]{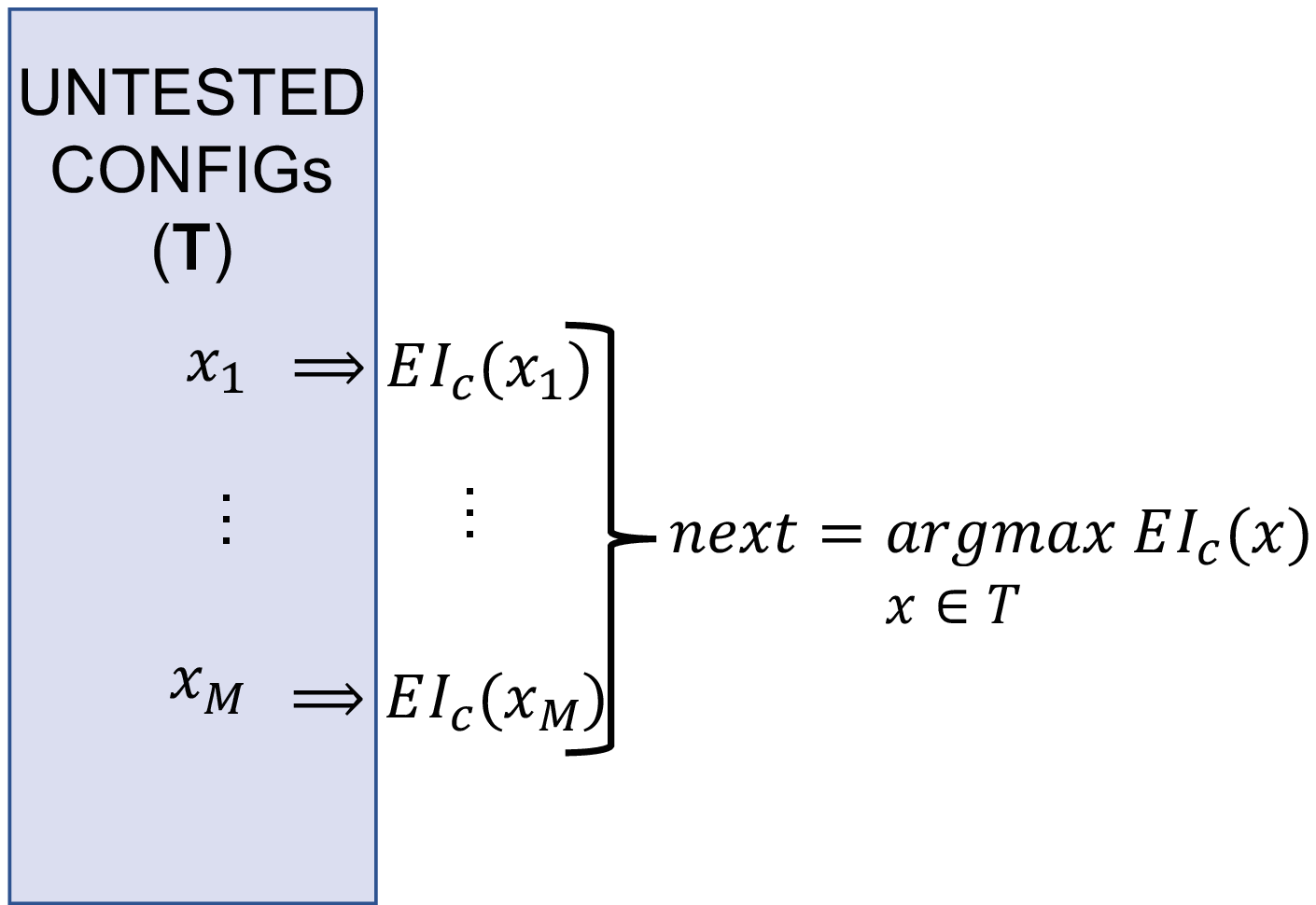}
        \caption{Typical BO approaches.}
        \label{fig:bo}
    \end{subfigure}
    \hfill
    \begin{subfigure}[h]{0.64\textwidth}
    \centering
       \includegraphics[scale=0.3]{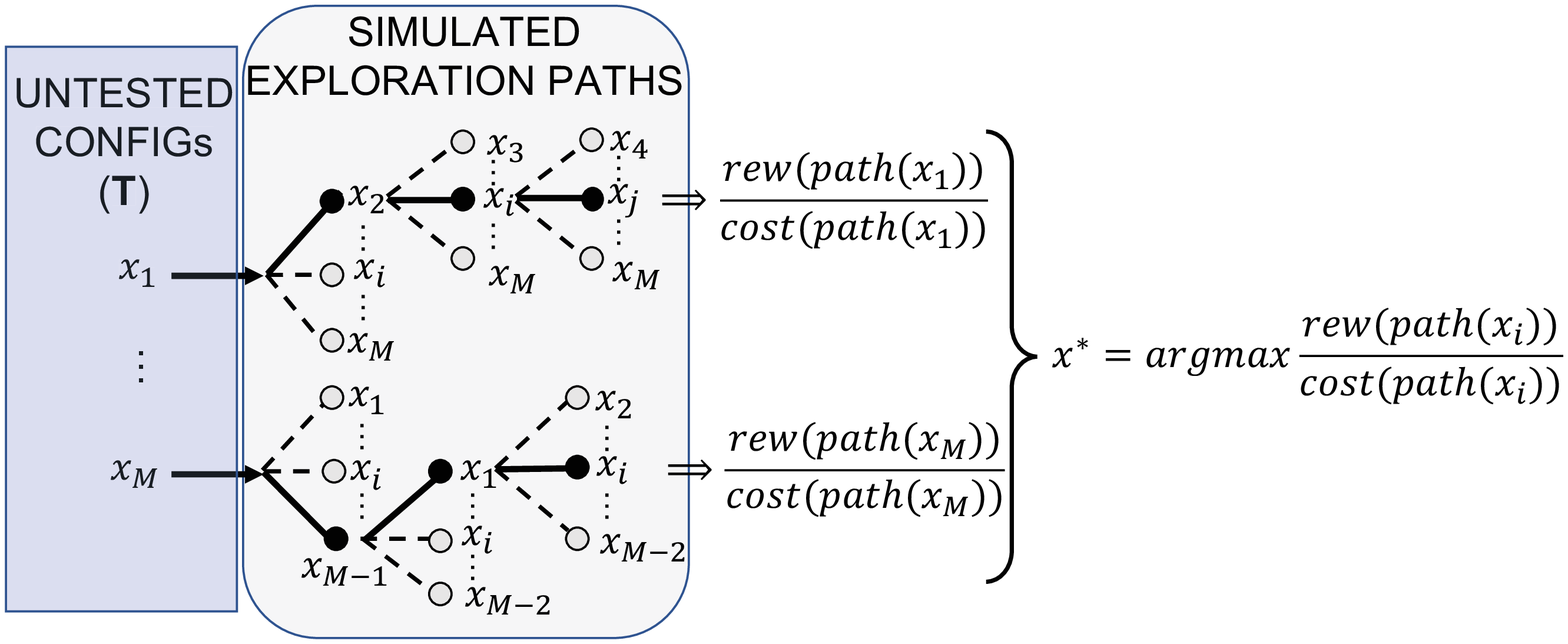}
        \caption{Lynceus.}
        \label{fig:lyn}
    \end{subfigure}
\caption{Selecting the next configuration in: (a) BO approaches; (b) \ts{}.  BO approaches maximize a one-step acquisition function, $EI_c$ in this case, that estimates the reward of sampling the \textit{next} configuration. \ts{}, instead, speculates on different exploration paths, by \textit{simulating} exploration paths. Dark/empty circles indicate the configurations selected/discarded at each step. For each configuration $x$, \ts{} computes the expected reward-to-cost ratio of the path that starts with $x$. Finally, it suggests the configuration $x^*$ that maximizes the long-term expected reward-to-cost ratio.}\label{fig:next_config}
\end{figure*}

\mynewpar{Acquisition function.} 
Given the current model of $f$, the acquisition function determines which point to evaluate next, among the set of points that are not yet in $S$.  
The acquisition function used by \ts{} is based on the {\em constrained expected improvement} ($EI_c$)~\cite{Gardner:2014}.
The $EI_c(x)$ for configuration $x$ is computed as the product of the probability that $x$ respects a given constraint, noted $P_C(x)$, and  the Expected Improvement of $x$, noted $EI(x)$. 
As its name suggests, $EI(x)$ estimates by how much configuration $x$ is expected to improve over the currently known optimum. Such expectation is computed taking into account both the expected value of $f(x)$ as predicted by the model, as well as the \textit{uncertainty} of the model on this prediction.
$EI(x)$ can be computed in closed form, assuming that $f$ follows a normal distribution~\cite{Jones:1998}. 
Specifically, $EI(x) =\big(y^* - \mu(x) \big) \Phi(z) + \sigma(x) \phi(z)$, where $\mu(x)$, resp. $\sigma(x)$, is the mean, resp. variance, of the prediction of the model of $x$; $\Phi$, resp., $\phi$, is the pdf, resp., CDF, of a standard normal distribution; and $z = (y^* - \mu(x)) /  \sigma(x)$.

In \ts{}, $y^*$ is the cost of the cheapest configuration profiled so far such that running a job takes at most $T_{max}$ time.
If there is no such configuration, $y^*$ is estimated as the
cost of the most expensive
configuration in $S$ plus three times the maximum standard deviation over the predictions on the points not in $S$~\cite{Lam:2017}.

$P_C(x)$ can be computed by training a regression algorithm on the target constraint variable, whose value is known for each point in $S$.
In \ts{}, $P_C(x) = P(T(x) \leq T_{max})$.
Instead of training a separate model for $T(x)$, \ts{} reuses the model that it already builds for $C(x)$, by leveraging the fact that $C(x) = T(x) \cdot U(x)$, where $U(x)$ is known. As such, rather than computing $P(T(x) \leq T_{max})$, \ts{} computes $P(C(x) \leq T_{max} \cdot U(x))$.

At each iteration, BO samples the configuration $x \notin S$ that maximizes $EI_c(x)$. $EI_c(x)$ has a  high value not only if $x$ is predicted --on average-- to be a good point, but also  if the uncertainty on $y(x)$ is high. This allows for   balancing \textit{exploitation} (testing points that are considered good) and \textit{exploration} (testing uncertain points) with the goal of improving the models' quality.

\mynewpar{Regression model.}
Computing $EI_c$  in closed form requires a regression model that assigns to each point $x$ a cdf $p(x)$ that is normally distributed $N(\mu(x), \sigma(x))$. To meet this requirement, \ts{} uses a {\em bagging ensemble}~\cite{Breiman:1996} of decision trees, i.e., a set of decision trees, each trained over a uniform random sub-set of $S$\footnote{Note that \ts{} can also operate using Gaussian Processes, as done by other BO approaches~\cite{Jones:1998,Brochu:2010}. We opted for a bagging ensemble of learners, since it offers more flexibility in the choice of the base learners to use.}. Then, \ts{} obtains $\mu(x)$ and $\sigma(x)$  based on the output of the individual predictors evaluated at $x$.  \ts{} uses these values to compute the $EI_c(x)$, assuming that the $p(x)$ associated with the ensemble of learners is normally distributed~\cite{Hutter:2011,Thornton:2013}.

\mynewpar{Stopping criterion.} Typical BO-based systems~\cite{Zeng:18,Alipourfard:2017,Didona:2016} stop the exploration phase once they detect that only marginal improvements are predicted by the model, e.g., when the $EI_c$ falls below 1\% for all unexplored configurations. \ts{} supports this classic stopping criterion and complements it to keep into account user defined constraints on the maximum budget available for the exploration phase.

\section{\ts{}}
\label{sec:lyn}

\ts{} takes as input the budget $B$, the maximum job runtime $T_{max}$ and a set $H$ of possible configurations. \ts{} then proceeds in an iterative fashion, similarly to typical BO approaches. At each iteration, \ts{} indicates a new configuration on which to profile the job. Once the job completes, the corresponding cost and performance information are used to update the regression model. The budget is also reduced by the cost incurred to run the job on the configuration. \ts{} stops when there are no more configurations to try with the available budget, or if the $EI_c$ for the unexplored configurations is marginal (below 1\%). 
The configuration recommended in the end is the one, among those sampled by \ts{}, with the lowest cost and with runtime within $T_{max}$.

\begin{figure*}[th!]
  \begin{subfigure}[h]{0.49\textwidth}
       \centering
       \includegraphics[scale = 0.35]{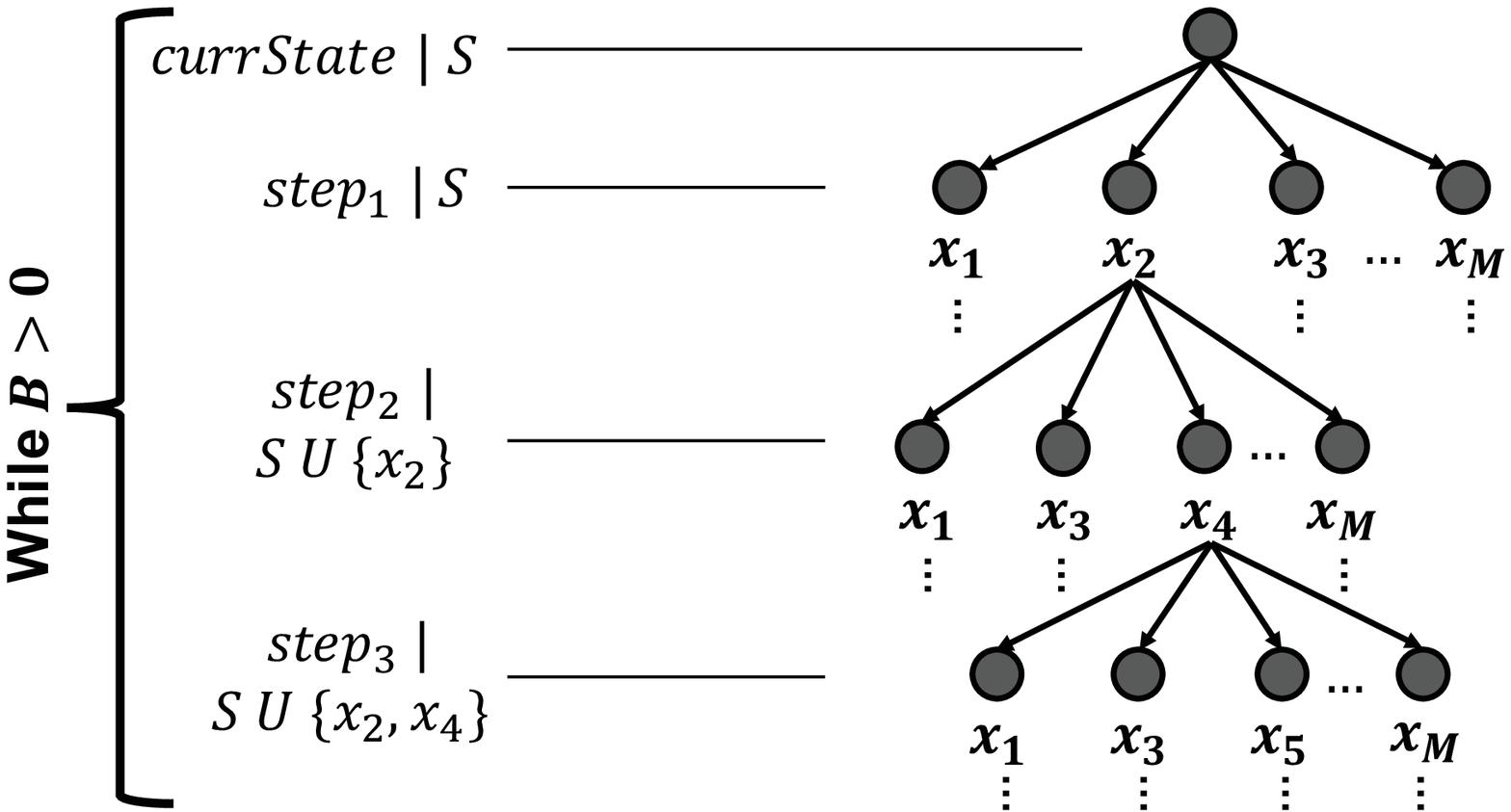}
        \caption{Intractable exhaustive formulation.}
        \label{fig:paths:exhaustive}
    \end{subfigure}
    \hfill
    \hfill
    \begin{subfigure}[h]{0.49\textwidth}
    \centering
       \includegraphics[scale=0.35]{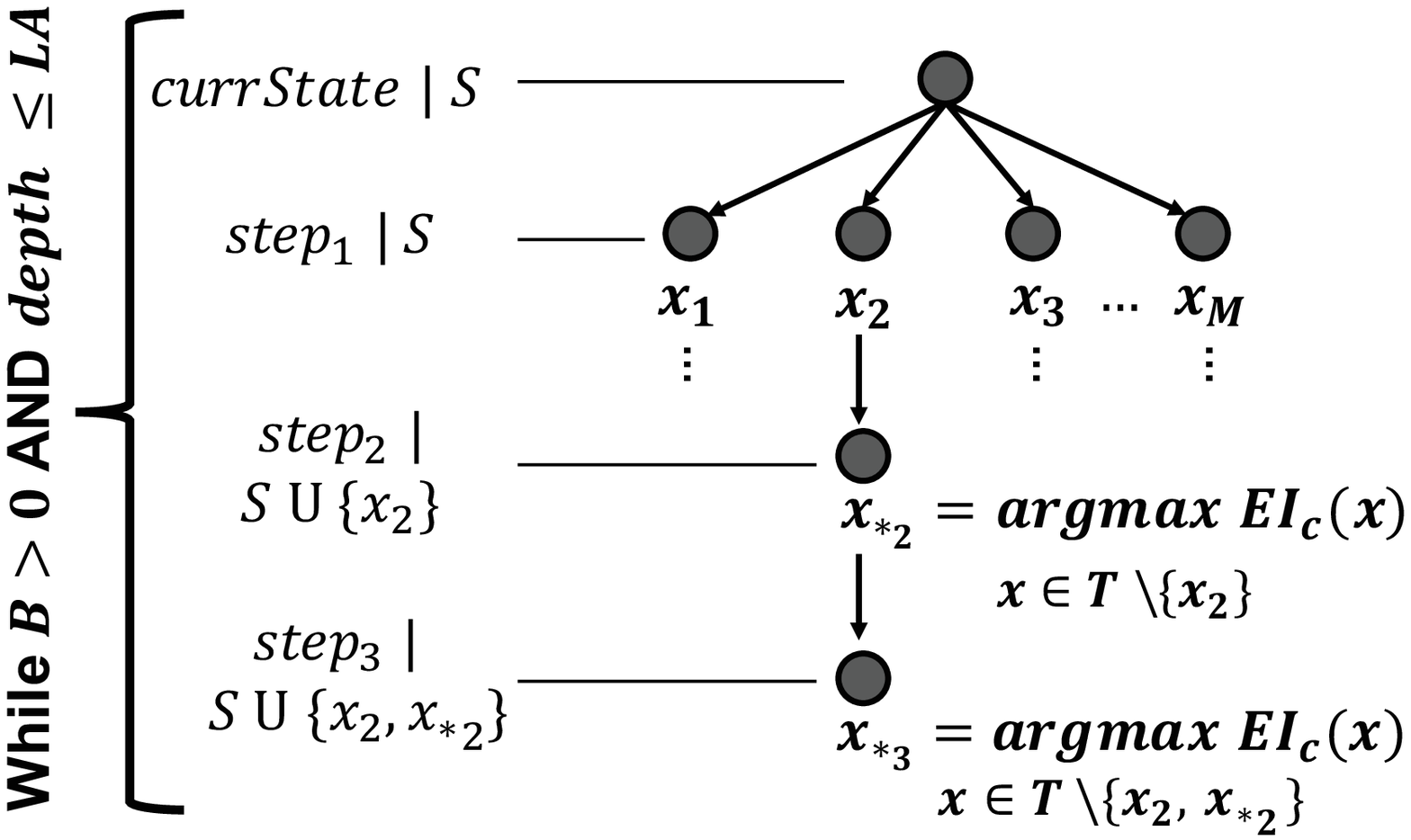}
        \caption{Approximate formulation employed by Lynceus.}
        \label{fig:paths:approx}
    \end{subfigure}
    \caption{(a) exhaustive approach that  enumerates the entire set of $M!$~exploration paths, given a set of $M$ unexplored configurations. (b)  The  search heuristics of \ts{s}: (i) an in-breadth search, at step 1, spanning all untested configurations; (ii) an in-depth search, at step $i>1$, which uses a BO-inspired principle and selects the configuration that maximizes the $EI_c$.
    The figure omits the technique used for incorporating the outcome of previous simulations in the model.}
    \label{fig:paths}
\end{figure*}

\subsection{Determining which configurations to sample}
\label{sec:lyn:which}
Figure~\ref{fig:next_config} provides an  overview of the selection process in typical BO approaches (Figure~\ref{fig:bo}) and in \ts{} (Figure~\ref{fig:lyn}). 
At each iteration of the optimization process, \ts{} speculates about several {\em exploration paths}. Each path corresponds to a possible sequence of configurations to be explored. To select the best path, the outcomes of testing the configurations in each path are simulated using a black-box model. These simulation results are then used to compute the \textit{reward} and the \textit{cost} of each path.

The \textit{reward} of a path corresponds, intuitively, to the aggregate reward resulting from exploring all the configurations in that path. The reward of a single configuration is given by its EIc, that is, the  cost improvement brought by that configuration, as predicted by the model, over the best configuration found so far.
The \textit{cost} of a path captures the predicted budget required to sample all configurations of the path.
Finally, \ts{} explores  the \textit{first} configuration of the path with the best reward/cost ratio. This approach  renders the optimization process of \ts{}  long-sighted and budget-aware.

\noindent \textit{Long-sighted:} 
By analyzing in foresight a \textit{sequence} of exploration steps (using a bounded look-ahead horizon) \ts{} defines  effective exploration policies, which can intentionally sacrifice the immediate reward stemming from the next exploration  in order to maximize the reward in the long term. This contrasts with existing BO approaches~\cite{Lam:2016,Lam:2017,Gonzalez:2016}, which use a greedy policy that maximizes a one-step/myopic acquisition function (such as $EI$). 

\noindent  \textit{Budget-aware:}  \ts{}  dynamically adjusts its ``explorative'' nature depending on the budget currently available. Compared to conventional BO schemes, \ts{} tends to favor the exploration of uncertain configurations, provided that this does not compromise the  budget available for future, less ``risky'' explorations. As a result, \ts{} adopts more explorative policies in the initial phase of the optimization process, when the model still has limited knowledge on the actual cost function and is, thus, more error prone. As the exploration progresses and the available budget diminishes, \ts{} tends to use a more risk-averse approach and to exploit the model's knowledge to maximize shorter term rewards.

\mynewpar{Challenges and solutions.}
Designing long-sighted optimization schemes, such as the one employed by \ts{}, requires tackling  two main challenges.
The first is related to the fact that the number of distinct exploration paths grows factorially with the unexplored configurations. As such, an idealized exhaustive approach, that analyzes all distinct exploration paths (illustrated in Figure~\ref{fig:paths:exhaustive}) would incur prohibitive computational costs in practical settings, forcing the use of approximations, i.e., search heuristics.
The second challenge is tied to the simulation of the outcomes of exploration steps at depth $i>1$. Such a simulation requires incorporating in the model used at step $i$ the effects of performing all previous explorations at steps $j<i$. However, configuration $x$ at step $j$ was not actually tested, but only simulated via a (Gaussian) black-box model that associates a non-null probability to \textit{any} possible cost value of $x$.
To ensure that the effects of exploring configuration $x$ at step $j$ are taken into account in the model used at step $j+1$, it would be necessary to marginalize over all possible cost values, and corresponding probabilities, predicted for $x$ by the model at step $j$.
Unfortunately, the closed form solution of such a nested marginalization problem implies prohibitive computational costs~\cite{Osborne09:gaussianprocesses} even for two-steps look-ahead. Thus, approximation techniques are required to make the problem tractable.

\ts{} tackles the above challenges  by means of three approximations, which ensure its scalability and viability. 

\noindent\textit{1)} The exploration paths considered by \ts{} are generated using a search heuristic that aims to balance the computational complexity of the optimization process and the effectiveness of the resulting exploration policy. This is achieved by using, in the first step, a breadth search policy that considers all untested configurations. At any subsequent step, instead, \ts{} employs a depth-first approach that selects the configuration that maximizes the $EI_c$, based on the current model's state. 
This BO-inspired heuristic allows for pruning significantly the search space, as it avoids that a path branches to consider all cases corresponding to choosing each possible configuration for the next step (except in the first).

\noindent\textit{2)} The in-depth simulation of a path is limited by a look-ahead window of size $LA$. Namely, the maximum length of an exploration path is limited to at most $LA$ steps, in addition to the first one. A path can be shorter than $LA$ steps in case the budget is depleted before reaching the $LA$-th step. If $LA$ is 0, \ts{} collapses to the traditional BO approach, where a single-step reward is maximized. Figure~\ref{fig:paths:approx} illustrates  the combined use of these two heuristics.

\noindent\textit{3)} 
To make the problem of simulating exploration paths  mathematically tractable, \ts{} discretizes the cost distribution output by the black-box model using the Gaussian-Hermite (G-H) quadrature~\cite[Chapter~5.3]{Gil:2007}. The G-H quadrature is used to approximate the value of integrals of the form $f(x) e^{-x^2}$ (such as the normal distribution that \ts{} associates with the outputs of its bagging regression model). The G-H quadrature produces $K$ $\langle$value, weight$\rangle$ pairs associated with the approximated function. In \ts{}, each value is a cost, and each weight captures, roughly speaking, the likelihood of the corresponding cost.

With these approximations, \ts{} simulates only $M$ paths ($M$ being the number of unexplored configurations) of length $LA$. Thus, \ts{}' complexity is $\mathcal{O}(K^{LA})$ since the G-H quadrature yields $K$ sub-trees at each step, up to depth $LA$.

\subsection{Sampling of sub-optimal configurations.} 
\label{sec:lyn:suboptimal}

Inaccuracies in the model can lead to exploring sub-optimal configurations, whose sampling can take a significant amount of resources (both cost and time). Prior works in the literature on BO~\cite{Duan:2009,Golovin:2017} suggest coping with this issue by simply aborting the exploration of configurations that are found to be worse than the best configuration identified so far. While such a simplistic approach does allow for saving resources, it is easy to see that it also suffers from a major drawback: since the cost of the configuration is unknown (because the testing was terminated prematurely) no feedback is given to the model and, as such, the (economical and temporal) resources spent prior to canceling its sampling are wasted in vain.

\mynewpar{Challenges and solutions.} Addressing the previous limitation requires tackling two tightly intertwined challenges.

The first challenge is related to \textit{how} to predict the cost of a configuration whose sampling has been timed out. Given the vast heterogeneity of modern cloud applications, we argue that, to maximize its interoperability, the technique  used to perform this prediction should be fully generic and transparent, i.e., it should impose no requirements or make no assumption on the underlying application. This excludes, for instance, designs that require the application to externalize periodic information of its progress rate or of its expected termination time.

The second challenge is related to \textit{when} to time out the sampling of a configuration: the later this is done, the more accurate the prediction can be (being fully accurate if, as an extreme, the sampling is not timed out at all), but also the smaller the gain in terms of  money (and time) saved due to cancelling the configuration's sampling. iTuned~\cite{Duan:2009}, for instance, takes a rather conservative approach and times out the sampling of a configuration once it has been in execution for twice as long as the fastest configuration found so far. Vizier~\cite{Golovin:2017}, instead, adopts a more aggressive (and non-transparent) approach based on monitoring the progress rate of the application in the configuration under test, say $x$, and comparing it with prior runs in different configurations. If after $t$ time units, the progress rate in $x$ turns out to be slower than the median of the progress rate at time $t$ for all the configurations tested so far, $x$ is deemed sub-optimal and its exploration is cancelled. Note that, unlike \ts{}, neither iTuned nor Vizier provide any information on the quality of timed out configurations. Hence, the choice of when to time out the sampling of a configuration only affects the amount of resources that are spent.
For \ts{}, the amount of information that the model gains about sampled configurations and its accuracy both depend on the timeout instant.

Compared to prior works, \ts{} addresses the latter problem by seeking a different trade-off regarding when to time out the sampling of a configuration $x$. Whenever the cost of $x$ is found to exceed the cost, noted $C(x^*)$, of the cheapest solution found so far (which must meet the user defined performance constraints), \ts{} times out the exploration. Since in the cloud the cost per unit of time of configuration $x$, $U(x)$, is known \textit{a priori}, it follows that the time out of the sampling of $x$ will occur at time $t=C(x^*)/U(x)$. At this time, \ts{} knows that configuration $x$ is not optimal.

At this  point \ts{} relies on the black-box model it uses to estimate the cost of unknown configurations
in order to predict the cost of the timed out configuration $x$. More precisely, the output distribution of the cost model, which follows a Gaussian distribution $N(\mu(x),\sigma(x))$, is conditioned to be strictly larger than $C(x^*)$. This yields a truncated Gaussian distribution~\cite{burkardt2014truncated,barr1999mean}, whose expected value can be computed in closed form as:
$$E(C(x)~|~C(x)>C(x^{*}))=\mu(x)+\sigma(x)\cdot\lambda(\alpha)$$
where $\lambda(\alpha)=\frac{\phi(\alpha)}{1-\Phi(\alpha)}$, 
$\alpha=\frac{C(x^{*}) - \mu(x)}{\sigma(x)}$ and $\phi$/$\Phi$ denote the pdf/CDF of a standard normal distribution, respectively. In \ts{} we use the expected value of this truncated Gaussian distribution, which is guaranteed to be larger than the cost of the current optimum, $C(x^*)$, to estimate the cost of $x$ and feed this information back to the model. 
This allows Lynceus to, unlike previous work, leverage the knowledge attained from suboptimal explorations which are timed-out earlier to prevent resource exhaustion to update its knowledge base.

\subsection{Detailed optimization algorithm}
\label{sec:lyn:detailed}

We first describe the state that \ts{} maintains and updates at each iteration. Then we describe the main optimization loop and detail how \ts{} speculates about the different exploration paths. Finally, we discuss how \ts{} copes with configurations whose exploration is revealed to be sub-optimal.

\mynewpar{State.} \ts{} maintains a state $\Sigma = \langle S, T, \beta, \chi\rangle$. $S$ is the current training set; $T$ is the set of unexplored configurations; $\beta$ is the remaining budget; and $\chi$ is the configuration currently deployed. 
\ts{} also associates a state with each step of each exploration path, to simulate how the optimization process would progress under different outcomes of the exploration of the untested configurations.

\mynewpar{Optimization loop.} Algorithm~\ref{alg:main} describes \ts{}'  optimization loop. The state is initialized as follows (Lines~\ref{alg:main:init:start}--~\ref{alg:main:init:end}): $S$ is empty; $T$ includes the whole set of  configurations; $\beta$ is set to $B$; and $\chi$ is set to $\bot$, as no configuration is currently deployed.

Then, \ts{} bootstraps the optimization loop (Lines~\ref{alg:main:bootstrap:start}--~\ref{alg:main:bootstrap:end}). \ts{} draws $N$ configurations at random\footnote{\ts{} uses Latin Hypercube Sampling~\cite{lhs}, a randomized technique to sample a multi-dimensional space that improves over random sampling.} and profiles the job with them.
Every time a job is run with a configuration $x$, \ts{} invokes the \texttt{Update} function. This function deploys the target configuration, runs the job and updates \ts{}' state. Namely, the budget is decreased by the amount of money needed to run the job, $C(x)$; a new pair $(x, C(x))$ is added to $S$; $x$ is removed from $T$; and the current configuration is set to $x$.

After the bootstrap phase, \ts{} enters the main loop (Lines~\ref{alg:main:loop:start}--\ref{alg:main:loop:end}). \ts{} decides the configuration $x$ to run next using the function \texttt{NextConfig}, executes the job on $x$ via the \texttt{Run} function, and updates its own state accordingly. Note that the \texttt{Run} function encapsulates the time out logic presented earlier (Section~\ref{sec:lyn:suboptimal}). 
The loop terminates when \texttt{NextConfig} returns a null value, meaning that there is no configuration that can be tried given the remaining budget or that the reward of every exploration path is marginal.

The \texttt{NextConfig} function operates as follows. It first identifies the set of configurations for which the estimated cost complies with the current budget. To this end, \ts{} queries the regression model to know which configurations are estimated to run the job with a cost lower than $\beta$ with a probability of at least $0.99$.  Then, for each of the viable configurations, the function computes the expected reward and the expected cost by means of the \texttt{ExplorePaths} function. Finally, \texttt{NextConfig} returns the configuration with the best reward to cost ratio. 

Note that the simulation of  exploration paths rooted at different untested configurations are independent problems that can be (and in our implementation are) solved in parallel. 

\begin{algorithm}[t!]
\scriptsize
\caption{Optimization loop}
\label{alg:main}
\begin{algorithmic}[1]
\Function{Main}{}
\State $\Sigma.\chi \lam \bot$\label{alg:main:init:start} \Comment{Config. currently deployed}
\State $\Sigma.S \lam \emptyset$ \Comment{Current training set}
\State $\Sigma.T \lam$ Whole configuration space \Comment{Set of untested configurations}
\State $\Sigma.\beta \lam B$ \Comment{Current budget}\label{alg:main:init:end}
\For{($i=0; i<N;i++$)}\label{alg:main:bootstrap:start}\Comment{Bootstrap}
\State $x \lam \textsl{LHC-sampling} (\Sigma.T)$\Comment{Select a random config.~using LHC-sampling}
\State $\textsc{Update} (\Sigma, x)$\Comment{Test config $x$ and update the state $\Sigma$}
\EndFor\label{alg:main:bootstrap:end}

\While{true}\label{alg:main:loop:start}
\State $x\lam \textsc{NextConfig} (\Sigma,LA)$ \Comment{Determine the next config to try}
\If{($x == null$)} \Comment{Stop exploration}
\State \Return $\textsl{argmin}_{y_x} \{ \Sigma.S \}$\Comment{Return best config tried}
\Else\Comment{Update model and state}
\State $\textsc{Update} (\Sigma, x)$ \Comment{Test config $x$ and update the state $\Sigma$}
\EndIf
\EndWhile\label{alg:main:loop:end}
\EndFunction

\Statex

\Function{Update}{$\Sigma$,$x$}
\State $\textsc{Deploy}(\Sigma.\chi, x)$\Comment{Set up the new config $x$, starting from the current config $\chi$}
\State $c \lam \textsc{Run}(x)$\Comment{Run the job on $x$ and return the cost}
\State $\Sigma.S \lam \Sigma.S \cup \{x, c\}$ \Comment{Add $(x,c)$ to the training set}
\State $\Sigma.T \lam \Sigma.T \setminus x$ \Comment{Remove $x$ from the set of untested configs}
\State $\Sigma.\chi \lam x $\Comment{Update config currently deployed}
\State $\Sigma.\beta \lam \beta - c$ \Comment{Decrease budget}
\EndFunction

\Statex

\Function{NextConfig}{$\Sigma$, LA}
\LineComment{Exclude configs prone to exceed the current available budget}
\State $\Gamma \lam \{x \in \Sigma.T : P(c(x)  \leq \Sigma.\beta | \Sigma.S) \geq 0.99 \}$ \label{alg:gamma}
\If{($\Gamma == \emptyset$)} \Comment{Stop exploration if all configs exceed budget}
\State \Return (null, 0) 
\Else \Comment{Compute rewards of exploration paths that start with any $x\in \Gamma$}
\State $\forall x \in \Gamma : (R_x, C_x) = \textsc{ExplorePaths}(\Sigma, x, LA)$
\State $sel \lam \textsl{argmax}_{x \in \Gamma}\{R_x/C_x\}$
\Comment{Select 1st config of path with max.~reward/cost}
\If{($R_{sel}<1\%$)} \Comment{Stop exploration if reward is marginal}
\State \Return null
\EndIf
\State \Return $sel$ \Comment{Else return selected config}
\EndIf
\EndFunction

\end{algorithmic}
\end{algorithm}

\begin{algorithm}[t]
\caption{\small Generate exploration paths starting from $x$}
\label{alg:rollout}
\begin{algorithmic}[1]
\scriptsize
\Function{ExplorePaths}{$\Sigma$, x, l}
\State $R \lam EI_c(x)$ \Comment{Set the reward of the path to the $EI_c$ of its first config} 
\State $C \lam Cost(x, S_k)$ \Comment{Set the cost of the path to the predicted cost of its first config}
\If{$l == 0$} \Comment{Look-ahead horizon was reached} 
\State \Return $(R, C)$ \Comment{Return the current path's reward and cost.}
\Else
\State $\langle c_i, w_i\rangle  \lam GH(f_c(x)), i=1,\ldots,N$ \Comment{Gauss-Hermite quadrature}
\For{$(i=1,\ldots,K$)}\Comment{Create state with speculated values}
\State $S' \lam \Sigma.S \cup \{(x, c_i)\}$ \Comment{Add config $x$ and speculated cost $c$  to training set}
\State $\beta' \lam \Sigma.\beta - c_i$ \Comment{Update available budget}
\State $T'\lam \Sigma.T \setminus \{x\}$ \Comment{Remove $x$ from the set of unexplored configs}
\State $\chi' \lam x$ \Comment{Set $x$ as the current config}
\State $\Sigma' =\{S', \beta', T', \chi'\}$ \Comment{State $\Sigma'$ includes the simulated cost ($c_i$) of $x$}
\State $x' \lam \textsc{NextStep}(\Sigma')$\Comment{Select next config $x'$ based on $\Sigma'$}
\If{$(x' == null)$}
\State continue \Comment{There's no suitable x'}
\EndIf
\LineComment{Compute reward and cost of sub-path of length $l-1$ rooted in $x'$}
\State $(r, c) \lam \textsc{ExplorePaths}(\Sigma',x', l-1)$ 
\State $C \lam C + w_i c  $  \Comment{Incorporate  cost\&reward of sub-path, weighted by $w_i$}
\State $R \lam  R + \gamma w_i r$ \Comment{Reward of future expl.steps is discounted by a factor $\gamma$}

\EndFor
\State \Return (R, C) \Comment{Return reward and cost of path of length $l$ rooted in  $x$}
\EndIf
\EndFunction

\Statex

\Function{NextStep}{$\Sigma$} \Comment{Select next config of expl. path at depth $i\ge 2$}
\LineComment{Exclude configs  prone to exceed the current available budget}
\State $\Gamma \lam \{x \in \Sigma.S : P(c(x)  \leq \Sigma.\beta | \Sigma.S) \geq 0.99\}$
\If{$\Gamma == \emptyset$} \Comment{Stop exploration if all configs exceed budget}
\State \Return (null, 0) 
\EndIf
\State \Return $\textsl{argmax}_{x \in \Gamma}\{EI_c(x)\}$ \Comment{Select config that maximizes $EI_c$}
\EndFunction
\end{algorithmic}
\end{algorithm}

\mynewpar{Exploration paths.} Algorithm~\ref{alg:rollout} provides the pseudo-code of the \texttt{ExplorePaths} function.  \texttt{ExplorePaths} takes as input the current state $\Sigma$ from which the path is starting, the configuration $x$ to explore in the current step, and the remaining length of the path $l$. 
Initially, when the function is called from within the main loop, $l$ is set to the value of the look-ahead window and is subsequently decremented every time \texttt{ExplorePaths} is invoked recursively.

\texttt{ExplorePaths} returns the expected reward and cost corresponding to using $x$ as the next step of the exploration path starting from state $\Sigma$. These values are given by the sum of two contributions: $i)$ the reward and cost corresponding to running the job on $x$; $ii)$ the weighted average of the rewards and costs of possible sub-paths that follow that exploration.

\texttt{ExplorePaths} operates as follows. First, it initializes the path's reward and cost with the (model's predicted) reward and cost of trying its first configuration $x$ (Lines 2--3).
The reward is computed as the $EI_c$ corresponding to $x$. The cost of the step is the mean cost of running the job on $x$ predicted by the black-box model. Then, the function generates the next steps for the path. If the remaining length of the look-ahead window is 0, then the path terminates. In this case, the reward and the value just computed are returned (Lines 3--6).
If $l > 0$, \texttt{ExplorePaths} generates the next steps of the path recursively. To this end, the function speculates about different possible costs $c_i$ associated with $x$, which are linked with likelihoods $w_i$ of being the real costs of running the job on $x$. The $\langle c_i, w_i\rangle$ pairs are obtained by computing the G-H quadrature on the p.d.f. that the black-box model predicts for the cost of $x$ (Line 7).

Each cost $c_i$ branches the path in a different sub-path in which the black-box model is updated with the speculated $\langle x, c_i\rangle$ configuration-cost pair, and in which the available budget is decreased by $c_i$. The augmented training set, the new budget and the updated set of untested configurations are encoded in a new state $\Sigma'$ (Lines 9--12).

The next configuration in the path is then computed by the \texttt{NextStep} function, which takes as input $\Sigma'$ (Lines 24--31). \texttt{NextStep} first computes the set of configurations that would not lead to a budget violation, if tested. 
If the set is empty, \texttt{NextStep} returns null. In this case, the path terminates, and \texttt{ExplorePaths} does not explore it further (Lines 14--16).
Else, \texttt{NextStep} returns the configuration $x'$ with the highest $EI_c$ in the set. In this case, \texttt{ExplorePaths} is invoked recursively to obtain the reward and cost values corresponding to following the sub-path that, from state $\Sigma'$, starts with $x'$ (Line 17). These values are used to update the reward and the cost corresponding to that path (Lines 18--19). 

When performing this update operation, the reward  values returned by \texttt{ExplorePaths} are multiplied by a {\em discount} factor $\gamma \in [0,1]$.
The lower the value of $\gamma$, the more \ts{} favors paths whose reward is higher in the early steps. If $\gamma = 0$, \ts{} discards any future rewards, and collapses to using the typical greedy BO algorithm. On the contrary,  if $\gamma = 1$, \ts{} gives the same weight to early and late rewards in the path. Our implementation uses $\gamma  = 0.9$, similarly to previous work~\cite{Lam:2016,Lam:2017}.

Finally, \texttt{ExplorePaths} returns the overall reward and the overall cost that one can expect if  $x$ is used as the next step in a path that starts from state $\Sigma$ (Line 21).

\section{Evaluation}\label{sec:eval}

Our evaluation addresses the following main questions: 
\begin{itemize}
    \item 
 By how much can \ts{} reduce the cost of the optimization process w.r.t~existing approaches (\S~\ref{sec:eval:quality})?
\item
 To what extent do the various features of \ts{} contribute to its effectiveness (\S~\ref{sec:eval:breakdown})?
\item How sensitive is \ts{}' performance to the technique used to time out sub-optimal configurations and to the LA setting (\S~\ref{sec:eval:timeout} and~\S~\ref{sec:eval:LA})?
\item What computational costs does \ts{} incur (\S~\ref{sec:eval:time})?
\end{itemize}
\subsection{Datasets}
\label{sec:exp:datasets}
We consider two datasets of heterogeneous data analytic jobs. The first dataset is composed of three Tensorflow jobs, which are characterized by a large configuration space defined over 5 dimensions. The second dataset is composed of several Hadoop and  Spark jobs that encompass smaller configuration spaces defined over 3 dimensions. These jobs have been used in the evaluation of the Scout~\cite{Hsu:2018b} and CherryPick~\cite{Alipourfard:2017} systems.

\mynewpar{Tensorflow Dataset.} We consider the distributed training of three  neural network models, i.e., CNN, RNN and Multilayer, over the MNIST dataset~\cite{Mnist}. The networks were implemented with   Tensorflow~\cite{Abadi:2016} using the parameter-server approach~\cite{Li:2014} and the ADAM optimizer~\cite{adam}.
The job terminates when the accuracy of the model reaches 0.85. We set a timeout of 10 minutes, after which a job is forcefully terminated. 
Table~\ref{tab:hyperandcloud} describes the tuning parameters that were considered. These include three hyper-parameters of the learning algorithm  and two cloud-related parameters, yielding  a total of 12 and 32 combinations of parameters, respectively.
We run our jobs on AWS EC2 and use 4 types of VMs. The VM clusters comprise 8 to 112 CPUs, for a total of 32 different cluster compositions. We did not  use GPUs to train these models, since CPUs are known to be more cost-efficient than GPUs when training neural models with the MNIST dataset~\cite{zhang2016efficient,wu2019comparative}. Table~\ref{tab:hyperandcloud} summarizes the cluster combinations that we use.
Overall, the configuration space for these jobs is composed of a total of $384$ configurations.

\begin{table}[t!]
\footnotesize
\begin{center}
\hspace*{-.25cm}\begin{tabular}{|c|c|c|c|}
\hline
\textbf{Parameter} & Learning rate & Batch size & Training mode\\ \hline
\textbf{Values} & $\{10^{-3}, 10^{-4}, 10^{-5}\}$ & \{16, 256\} & \{sync, async\}\\ \hline
\end{tabular}

\hspace*{-.25cm}\begin{tabular}{|l|c|c|}
\hline
\textbf{VM type} & \textbf{VM characteristics} & \textbf{\#VMs}                     \\ \hline
t2.small         & \{1 VCPU, 2 GB RAM\}        & \{8, 16, 32, 48, 64, 80, 96, 112\} \\ \hline
t2.medium        & \{2 VCPU, 4 GB RAM\}        & \{4, 8, 16, 24, 32, 40, 48, 56\}   \\ \hline
t2.xlarge        & \{4 VCPU, 16 GB RAM\}       & \{2, 4, 8, 12, 16, 20, 24, 28\}    \\ \hline
t2.2xlarge       & \{8 VCPU, 32 GB RAM\}       & \{1, 2, 4, 6, 8, 10, 12, 14\}      \\ \hline
\end{tabular}
\end{center}
\caption{TensorFlow parameters and cloud configurations.}
\label{tab:hyperandcloud}
\end{table}

\mynewpar{Scout and CherryPick datasets.} The Scout dataset~\cite{Hsu:2018} is composed of 18 Hadoop and Spark jobs of the HiBench~\cite{hibench} and spark-ref~\cite{spark-ref} benchmarks. The CherryPick dataset~\cite{Alipourfard:2017} is composed of 5 jobs: TPC-H~\cite{tpch}, TPC-DS~\cite{tpcds}, Terasort, Spark Kmeans~\cite{spark}, and Spark Regression~\cite{spark}. These jobs stress CPU, network and memory resources differently, hence allowing us to evaluate \ts{} in heterogeneous use cases.

Both sets of jobs were run on AWS EC2, using different sets of VM types, but not varying any application-level parameter. Overall, the Scout dataset considers a total of 69 different configurations, whereas the configuration space for the CherryPick dataset ranges from 47 to 72 points.
Additional details on the jobs can be found in the original papers. 

\subsection{Methodology}
\label{sec:exp:met}
\mynewpar{Compared systems.} 
We compare \ts{} with the traditional BO approach, used by state-of-the-art systems to optimize data analytic jobs, such as CherryPick~\cite{Alipourfard:2017} and Arrow~\cite{Hsu:2018b}. We refer to this approach as BO. We also consider a simple random approach (RND) to establish a baseline on the complexity of the optimization task.  We consider four values for the look-ahead parameter (LA) in \ts{}, i.e., LA=\{0, 1, 2, 3\}. LA=0 corresponds to a traditional BO approach, using as acquisition function the $EIc$ per dollar~\cite{Brochu:2010}, i.e., the ratio between the $EIc$ and the expected cost of a configuration. \ts{} and BO use a bagging ensemble of 10 random trees to build the cost model of the job, as in recent BO systems~\cite{Hutter:2011,Thornton:2013,Didona:2016}.

\begin{figure*}[t!]
  \begin{subfigure}[h]{0.33\textwidth}
       \includegraphics[scale = 0.55]{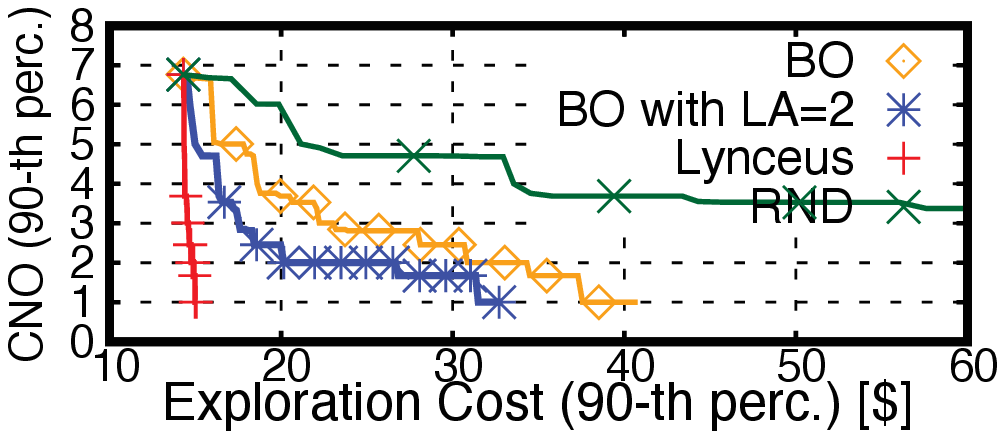}
        \caption{CNN.}
        \label{fig:cnn}
    \end{subfigure}
    \begin{subfigure}[h]{0.33\textwidth}
       \includegraphics[scale=0.55]{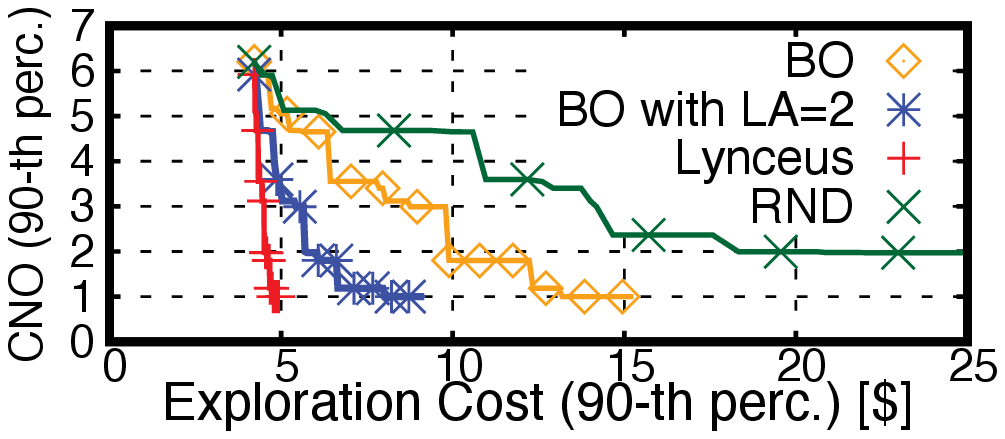}
        \caption{RNN.}
        \label{fig:rnn}
    \end{subfigure}
    \begin{subfigure}[h]{0.33\textwidth}
       \includegraphics[scale=0.55]{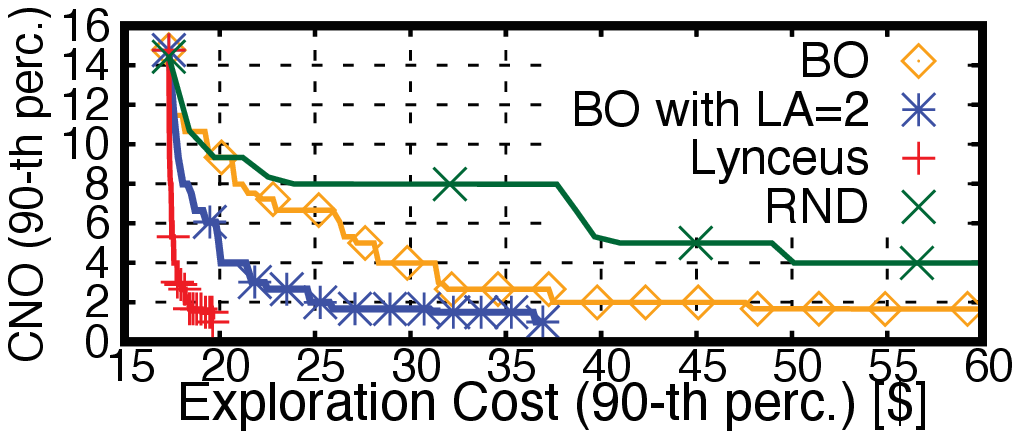}
        \caption{Multilayer.}
        \label{fig:mlp}
    \end{subfigure}
\caption{90-th percentile of the CNO achieved by \ts{}, BO with look-ahead 2, BO without look-ahead and RND for the Tensorflow jobs, as a function of the 90-th percentile of the exploration cost.
}
\label{fig:TF}
\end{figure*}

\mynewpar{Experiments.}
We perform our evaluation via a simulation approach, which uses the performance data of the Tensorflow, Scout and CherryPick datasets.
In each experiment we run an optimizer 100 times against a target job. Each run uses a different set of initial configurations to bootstrap the model. 

\mynewpar{Metrics.}
We use the Cost Normalized w.r.t. the Optimum (CNO) to evaluate the quality of the configurations recommended by an optimizer. Noting $x^*$ the optimal configuration, and $x$ the configuration suggested by an optimizer, the CNO achieved by the optimizer is $cost(x)/cost(x^*)$. Hence, the lower the CNO, the better. The optimal value for CNO is 1. In order to evaluate the cost-effectiveness of the optimizers we also measure the monetary cost consumed by each optimizer during the exploration phase (to deploy and sample a job in different cloud configurations).

\mynewpar{Budget.} To ensure a fair comparison with   the baselines considered in this study, which do not consider any  constraint on the exploration cost, we set the budget to infinity for \ts{}. By looking at how the budget is used over time it is possible to infer the behavior of \ts{} for different budget values.

\mynewpar{Default settings.} We set the initial number of samples, $N$, in a way that accounts for the size of the configuration space of each job. Specifically, noting with $\mathcal{C}$ the job's configuration space, we define $N$ as the max of (i) 3\% of the cardinality of $\mathcal{C}$ (a percentage also used in previous works~\cite{Didona:2016}) and (ii) the number of dimensions of $\mathcal{C}$. Unless stated otherwise, \ts{} uses LA=2 and the Truncated Gaussian timeout policy. We evaluate different timeout policies in Section~\ref{sec:eval:timeout} and lower values for LA in Section~\ref{sec:eval:LA}. We do not report results for larger LA values as, in our experiments, the gains deriving from setting LA=3 were marginal w.r.t. LA=2.
Finally, we set the time constraint for each job in such a way that it is satisfied by roughly half of the possible configurations.

\subsection{Cost of the optimization process}
\label{sec:eval:quality}

\mynewpar{Tensorflow jobs.}
We  evaluated the advantages of jointly optimizing  cloud and application parameters in Section~\ref{sec:challenges}. Thus, in the following, we assume, for fairness, that all the compared solutions are faced with the same configuration space that includes both cloud and application parameters.

Figure~\ref{fig:TF} reports the 90-th percentile of the CNO achieved by: \ts{}, BO with look-ahead (set to 2, as in \ts{}), BO without look-ahead and RND, as a function of the 90-th percentile of the exploration cost, for the three Tensorflow jobs.
The costs corresponding to the initial sampling phase are not explicitly shown, as they are the same for all approaches, but are taken into account and added to the first cost represented in each plot. \ts{} consistently outperforms the baseline approaches by reaching the optimal configuration at a lower exploration cost. In particular, for Multilayer (Fig.~\ref{fig:mlp}), while \ts{} reaches the optimal configuration after spending $\sim$\$19.5, the base BO technique requires spending $\sim$\$230, which corresponds to an improvement of more than $11\times$. For CNN (Fig.~\ref{fig:cnn}) and RNN (Fig.~\ref{fig:rnn}) the cost reduction to identify the optimum is lower, but still substantial, amounting to $\sim$2$\times$.

\begin{figure}[b!]
    \begin{subfigure}[h]{0.22\textwidth}
       \includegraphics[scale=0.55]{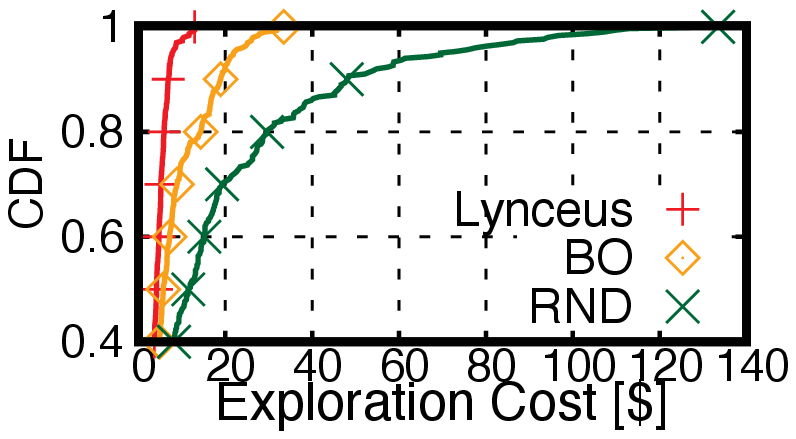}  
       \caption{CNO = 2}
       \label{fig:CNO2}
    \end{subfigure}
    \hspace{4pt}
  \begin{subfigure}[h]{0.22\textwidth}
      \centering
       \includegraphics[scale = 0.55]{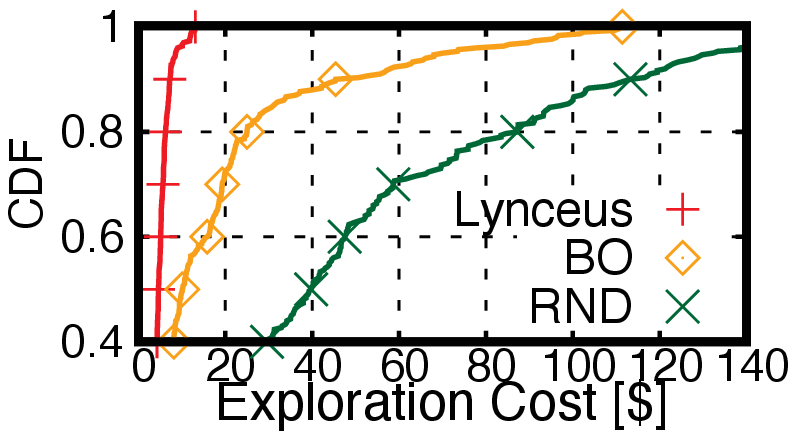} 
       \caption{CNO = 1.1}
       \label{fig:CNO10}
    \end{subfigure}
\caption{CDFs of the exploration cost for CNO=2 and CNO=1.1 (Tensorflow datasets).}
\label{fig:cdfs:TF}
\end{figure}

Figure~\ref{fig:cdfs:TF}  shows the CDF of the exploration cost to find a configuration that is either $2\times$ (Fig.~\ref{fig:CNO2}) or $10\%$ (Fig.~\ref{fig:CNO10})
away from the optimum, allowing us to assess the effectiveness of \ts{} in identifying configurations at different distances from the optimum. At the 90-th percentile, \ts{} spends $\sim$2.7$\times$ less than BO to identify configurations $2\times$ away from the optimum (Fig.~\ref{fig:CNO2}), with gains up to $\sim$6.2$\times$ when considering the more challenging problem of identifying configurations that are only 10\% worse than the optimum (Fig.~\ref{fig:CNO10}).

\begin{figure}[b!]
  \begin{subfigure}[h]{0.22\textwidth}
      \centering
       \includegraphics[scale = 0.55]{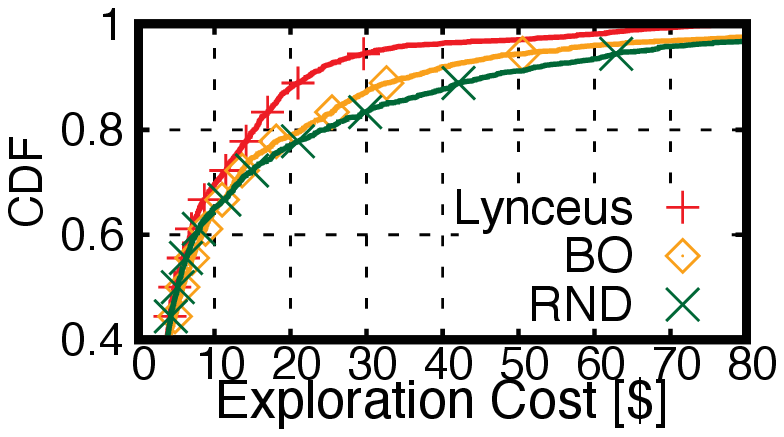}
        \caption{Scout jobs.}
       \label{fig:others:scout}
    \end{subfigure}
    \hspace{4pt}
    \begin{subfigure}[h]{0.22\textwidth}
       \includegraphics[scale=0.55]{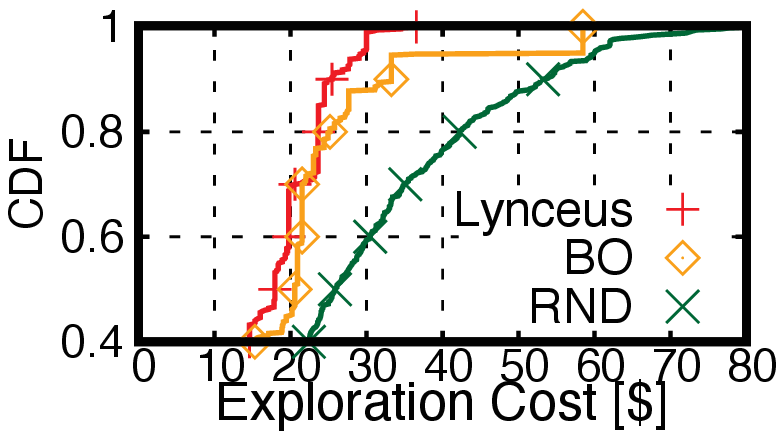}
       \caption{CherryPick jobs.}
       \label{fig:others:cp}
    \end{subfigure}
\caption{CDFs of the exploration cost for CNO = 1.1 (Scout and CherryPick datasets).}
\label{fig:cdfs_others}
\end{figure}

\mynewpar{Scout and CherryPick jobs.}
Figure~\ref{fig:cdfs_others} reports the CDFs of the exploration cost to identify configurations at 10\% from the optimum, for the Scout and CherryPick datasets. At the 90-th percentile, the gains of \ts{} over BO remain remarkable but less pronounced than for the Tensorflow datasets -- i.e., BO spends  $\sim$60\% and 
$\sim$48\% more 
for the Scout and Cherrypick datasets, respectively. This is due to the lower dimensionality of the search space (and thus to the lower complexity of the optimization problem), which decreases the benefits achievable by employing a more careful planning policy.

\begin{figure*}[t!]
  \begin{subfigure}[h]{0.33\textwidth}
       \includegraphics[scale = 0.56]{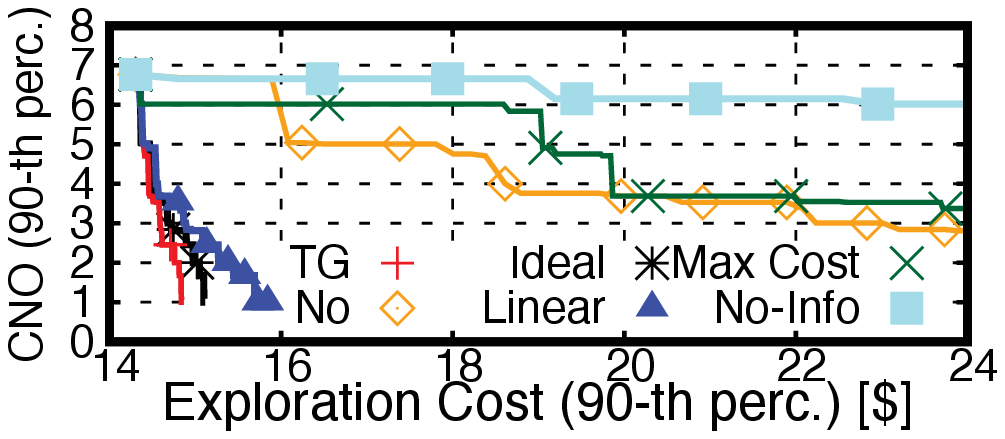}
        \caption{CNN.}
        \label{fig:timeout:cnn}
    \end{subfigure}
    \begin{subfigure}[h]{0.33\textwidth}
       \includegraphics[scale=0.56]{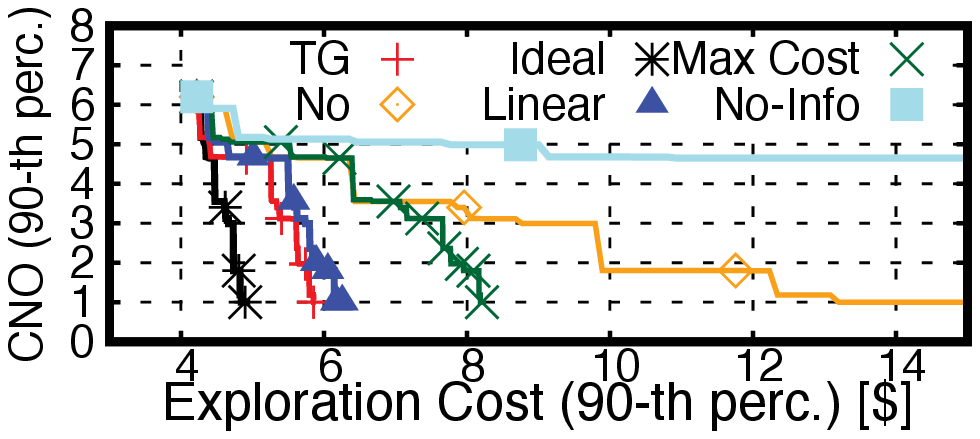}
        \caption{RNN.}
        \label{fig:timeout:rnn}
    \end{subfigure}
    \begin{subfigure}[h]{0.33\textwidth}
       \includegraphics[scale=0.56]{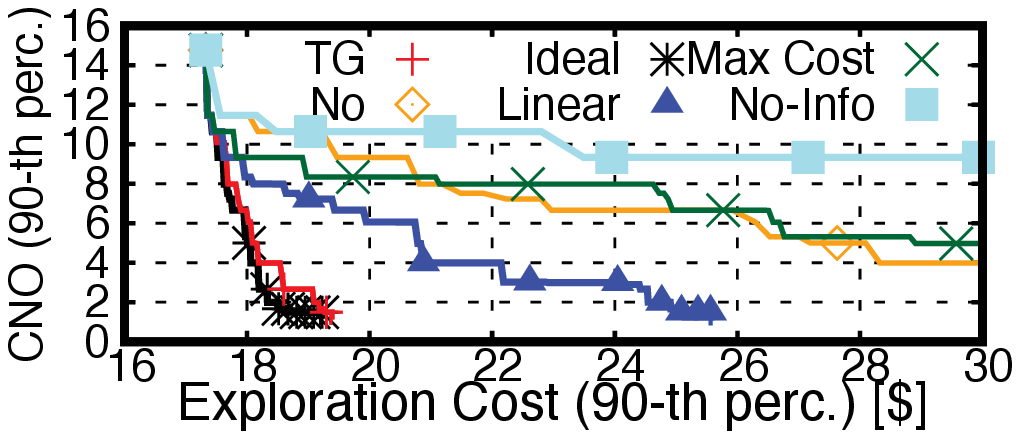}
        \caption{Multilayer.}
        \label{fig:timeout:mlp}
    \end{subfigure}
\caption{90-th percentile of the CNO achieved by several versions of the timeout as a function of the 90-th percentile of the exploration cost. \ts{}' method (TG) is the closest to an ideal approach that updates the  model  with exact values. }\label{fig:timeout}
\end{figure*}

\begin{figure*}[t!]
  \begin{subfigure}[h]{0.33\textwidth}
       \includegraphics[scale = 0.55]{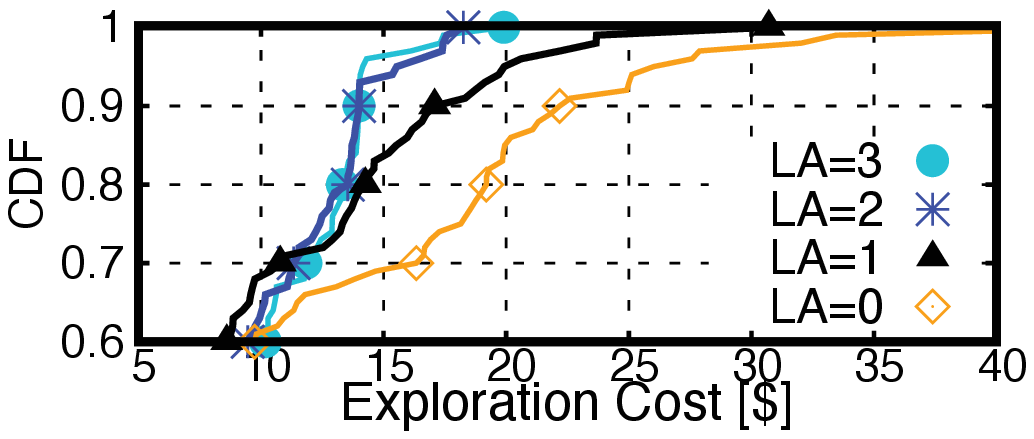}
        \caption{CNN.}
        \label{fig:break:cnn}
    \end{subfigure}
    \begin{subfigure}[h]{0.33\textwidth}
       \includegraphics[scale=0.55]{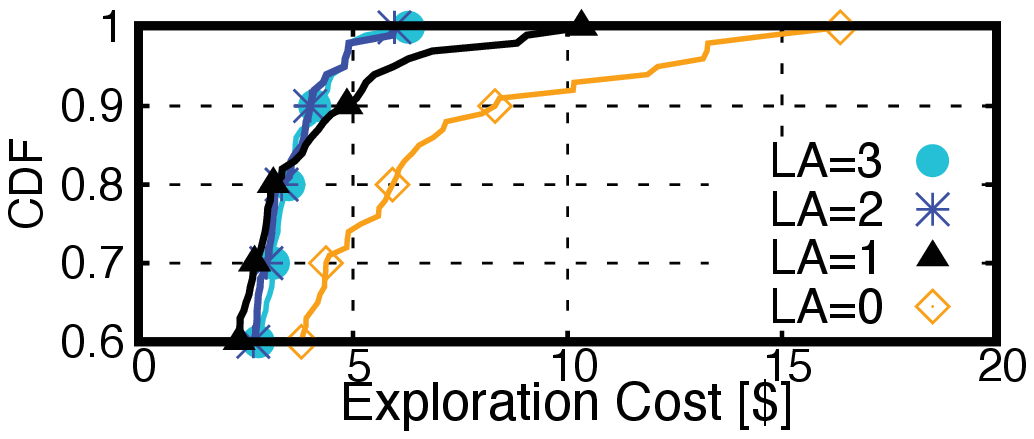}
        \caption{RNN.}
        \label{fig:break:rnn}
    \end{subfigure}
    \begin{subfigure}[h]{0.33\textwidth}
       \includegraphics[scale=0.55]{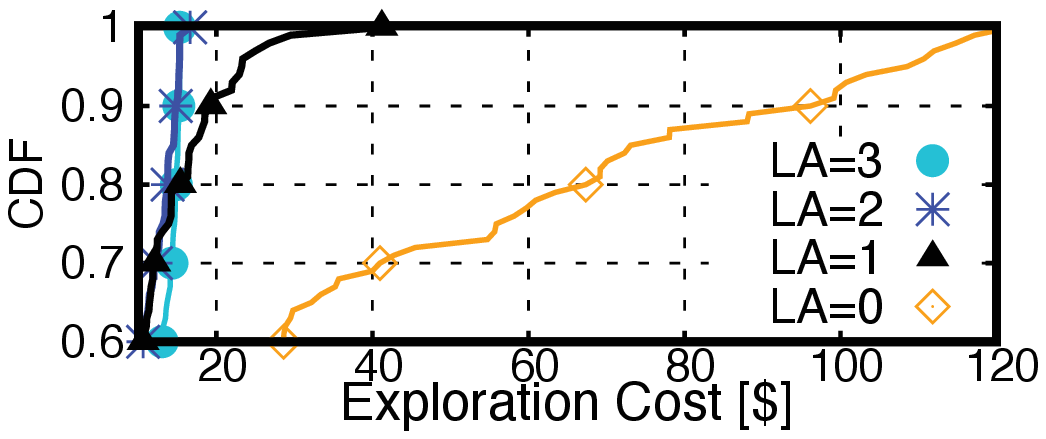}
        \caption{Multilayer.}
        \label{fig:break:mlp}
    \end{subfigure}
\caption{CDFs of the exploration cost achieved by \ts{} without the timeout feature and with LA=\{3, 2 (default value), 1, 0\}, for a CNO of 10\%. The use of look-ahead provides clear benefits, but the gains diminish using horizons beyond 2.}
\label{fig:break}
\end{figure*}

\subsection{Breakdown of the improvements}
\label{sec:eval:breakdown}

Let us now quantify the benefits deriving from the two key novel features of \ts{}:  i) the use of look-ahead to identify which configurations to sample,  and ii) the timeout mechanism when sampling sub-optimal configurations. To this end, let us return to Figure~\ref{fig:TF}. By comparing the base BO technique to the baseline using BO with LA=2, one can quantify the gains achieved using look-ahead. The benefits stemming from the timeout policy can instead be assessed by comparing \ts{} with the baseline of BO equipped with LA=2.

Overall, both mechanisms play an important role. While the use of look-ahead appears to be particularly relevant in the early stage of the optimization process, when the identified configurations are still relatively distant from the optimum, the timeout tends to provide the largest gains the closer \ts{} is to the optimum. For instance when the CNO is equal to 2, the use of look-ahead allows for achieving a cost reduction ranging from 40\% to 50\%, across all networks. From that point on, the benefits of look-ahead, although still significant in RNN and CNN, tend to become less relevant when compared to the ones stemming from the use of timeout. This can be explained considering that, in the early stage of the optimization, planning ahead which configurations to sample allows for exploring the configuration space in a more cost-effective way. Once configurations closer to the optimum are found, the use of timeout becomes extremely effective by imposing a strict upper bound on the cost of future explorations.

\subsection{Sensitivity to the timeout implementation}
\label{sec:eval:timeout}
Figure~\ref{fig:timeout} reports the 90-th percentile of the CNO, as a function of the 90-th percentile of the exploration cost when using various policies for estimating the full cost of running a job in a ``timed out'' configuration, namely: the Truncated Gaussian (TG) approach used by \ts{}; an approach (NO-INFO) similar to the one used in iTuned~\cite{Duan:2009} and Vizier~\cite{Golovin:2017}, where configurations are timed out if their cost exceeds by 2$\times$ the cost of the current optimum and where the model is provided with no information on timed out configurations; an Ideal approach that updates the model with the exact cost of fully executing the job; a Max Cost approach that feeds the model with the highest cost seen so far; a baseline that does not use the timeout feature (No); an approach (Linear) that assumes the availability of information on the job's progress and uses a linear model to estimate its full cost. For Linear, as we consider jobs for training neural networks, we use as progress indicator the accuracy reached by the model upon timeout and linearly extend it until the desired target accuracy (85\%). In order to evaluate the timeout feature in isolation, these results were obtained disabling the look-ahead.

The TG approach is consistently better than the others, getting very close to the Ideal timeout for the Multilayer and CNN networks (Figs.~\ref{fig:timeout:mlp} and~\ref{fig:timeout:cnn}). The Max cost and NO-INFO approaches are visibly worse than the others, showing that such simplistic approaches are clearly unfit and the relevance of feeding the model with  accurate information when a configuration is timed out. The shortcomings of the Linear approach are also evident for Multilayer, where it clearly is outperformed by TG. We argue that this is due to the inadequacy of a linear model in predicting the time remaining to achieve the target accuracy for the job. While this issue might be tackled using more complex (non-linear) models, these data show that such a complexity is, in practice, unnecessary, since the proposed TG approach is very close to the ideal approach.

\subsection{Sensitivity to the LA setting}
\label{sec:eval:LA}

Figure~\ref{fig:break} compares the CDFs of the exploration cost achieved by \ts{} (which uses LA=2) with the CDFs of three versions of \ts{} that use LA=\{3,1,0\}. The CDFs were obtained for a fixed CNO of 10\% and without using the timeout mechanism. 
The plots allow us to draw two main conclusions. 
First, even small look-ahead horizons can significantly enhance the efficiency of the exploration process. For instance, at the 90-th percentile,  LA=1 allows for achieving a five-fold reduction of the exploration cost versus a greedy approach (LA=0). 
Second, deeper look-ahead horizons tend to have diminishing gains,  becoming marginal beyond LA=2. This is unsurprising, since the deeper the look-ahead horizon, the larger the probability that the model-based simulation of future explorations is stale, yielding limited benefits.

\subsection{Prediction time}
\label{sec:eval:time}
Table~\ref{tab:time} reports the average time needed to predict the next configuration while varying the look-ahead's depth. 
In particular, the table refers to RNN, but we have obtained similar values for CNN and Multilayer, since the cardinality of the search space is the same. We report results for the Tensorflow jobs, which have the largest configuration space among the jobs we consider and, as such, impose the largest computational costs. The simulations are run on machines with Intel Xeon Gold 6138 CPU with 20 physical cores and 64 GB of main memory. As expected, \ts{}' prediction time grows with the length of the look-ahead window. With LA=2 (the default for \ts{}) the average computation time is around one second --- a latency that we argue to be perfectly affordable in the context of data analytic jobs.

\begin{table}[t]
\footnotesize
\centering
\begin{tabular}{|c|c|c|c|c|}
\hline
\textbf{Optimizer}             &   \begin{tabular}[c]{@{}c@{}}BO, Lyn\\ (LA=0)\end{tabular}  & \begin{tabular}[c]{@{}c@{}}Lyn\\ (LA=1)\end{tabular} &
\begin{tabular}[c]{@{}c@{}}Lyn\\ (LA=2)\end{tabular} &\begin{tabular}[c]{@{}c@{}}Lyn\\ (LA=3)\end{tabular} \\ \hline
\textbf{Avg seconds to next()} & 0.05 & 0.36 & 0.99 & 2.4 \\ \hline
\end{tabular}\caption{Average time needed to select the next config.}
\label{tab:time}
\end{table}

\section{Conclusions and future work}
\label{sec:conclusions}
We presented \ts{}, a new tool to provision and tune data analytic jobs. \ts{} implements a novel approach that combines cross-layer optimization, budget awareness, long-sightedness, and the ability to cancel sub-optimal sampling while still improving the model. 
\ts{} consistently outperforms  state-of-the-art approaches, identifying configurations that are up to $3.7\times$ cheaper --- thanks to the joint optimization of cloud and application parameters --- and reducing the cost of the optimization process by up to $11\times $ --- thanks to its novel optimization method.
As a final note, \ts{} can be extended to consider multiple constraints (e.g., one may want to enforce that the energy consumed to execute the job is also within a given threshold) and to take into account the costs associated with bootstrapping VMs during the exploration phase. An evaluation of these mechanisms is left for future work.



\bibliographystyle{IEEEtran}
\bibliography{main}

\end{document}